\documentclass{llncs}
\pdfoutput=1

\usepackage[pdftex]{graphicx}
\usepackage[english]{babel}
\usepackage{amsmath}
\usepackage{color}
\usepackage{amsfonts}
\usepackage{acronym}
\usepackage{array}
\usepackage{fixltx2e}
\usepackage{acronym}
\usepackage{listings}
\usepackage{url}
\usepackage[nocompress]{cite}
\usepackage{amssymb}
\usepackage{diagbox}
\usepackage{array}
\usepackage{subfig}
\usepackage{adjustbox}
\usepackage{pgf}
\usepackage{tikz}
\usepackage{courier}

\lstdefinelanguage{Modelica}
{
  morekeywords = {model,discrete,when,parameter,initial,equation,der,then,end,reinit,pre},
  sensitive = true,
  escapeinside={/+}{+/},
}
\lstnewenvironment{Modelica}
{\lstset{language={Modelica}
}} {}

\begin{document}

\title{Hybrid Simulation Safety:\\Limbos and Zero Crossings}
\titlerunning{Hybrid Simulation Safety: Limbos and Zero Crossings}
\author{David Broman}
\authorrunning{David Broman} 
\institute{KTH Royal Institute of Technology\\Sweden\\
\email{dbro@kth.se}}

\maketitle

\begin{abstract}
  Physical systems can be naturally modeled by combining continuous
  and discrete models. Such hybrid models may simplify the modeling
  task of complex system, as well as increase simulation
  performance. Moreover, modern simulation engines can often
  efficiently generate simulation traces, but how do we know that the
  simulation results are correct?  If we detect an error, is the error
  in the model or in the simulation itself? This paper discusses
  the problem of simulation safety, with the focus on hybrid modeling
  and simulation. In particular, two key aspects are studied: safe
  zero-crossing detection and deterministic hybrid event handling. The
  problems and solutions are discussed and partially implemented in
  Modelica and Ptolemy II.
\end{abstract}


\section{Introduction}
\label{sec:intro}

Modeling is a core activity both within science and engineering. In
various domains, there are different kinds of models, such as
dynamic models, probabilistic models, software models, and business
models. In general, a \emph{model} is an abstraction of something,
where this thing can be a process, a system, a behavior, or another
model. 

Both scientists and engineers make extensive use of models, but for
different reasons. As Lee~\cite{Lee:2016} points out, scientists
construct models \emph{to understand} the thing being modeled, whereas
engineers use models \emph{to construct} the thing being modeled. In both
cases, the abstraction (the model) contains fewer details than the
thing being modeled, which enables the possibility to \emph{analyze}
the model. Such analysis can include formal verification, statistical
analysis, or simulation.

The latter, \emph{simulation} of models, is the main topic of this
paper. Simulation can be seen as a way to perform experiments on a
model, instead of on the system or process being
modeled~\cite{Cellier:1991}. There are many reasons for using modeling and
simulation. It can be too
dangerous to perform experiments on real systems. It can be cheaper
to perform simulations, or the system being modeled might not yet
exist. 

Regardless of the reason for doing modeling and simulation, it is
vital to trust the simulation result to some degree. We say that the
\emph{fidelity} of the model is to what extent the model correctly
represents the thing being modeled. Lee~\cite{Lee:2014,Lee:2016} often
stresses the distinction between the model and what is being modeled,
by giving the famous quote by Golomb~\cite{Golomb:1971}: ``you will
never strike oil by drilling through the map''. High model fidelity is
necessary, but not sufficient to enable trust of simulation
results. To trust the map, as an example of a model, we also need to
interpret the map safely. For instance, if an English speaking
engineer is using a Russian map to find oil, even a map of high
fidelity can lead to incorrect conclusions. Misinterpretations of the
map (the model) can result in false positives (drilling through an oil
pipe instead of an oil field) or true negatives (drilling through a
mine field instead of an oil field). As a consequence, to trust the
use of models, not only high model fidelity is needed, but also safe
interpretation of the model.

If we make the analogy between a model and a computer program, we can
distinguish between two kinds of errors~\cite{Cardelli:2004}: i)
\emph{untrapped errors} that can go unnoticed and then later result in
arbitrary incorrect behavior, and ii) \emph{trapped errors} that are
handled directly or before they occur. For a computer program written
in the C programming language, an array out-of-bound error can lead to
memory corruption, where the actual problem can first go unnoticed,
and then crashes the system at a later point in time. This is an example
of an untrapped error. By contrast, an array out-of-bound error in
Java results in a Java exception, which happens directly when it
occurs, and makes it possible for the program itself to handle the
error. This latter case is an example of a trapped error. A program
language where all errors are trapped errors, either detected at
compile time using a type system, or at runtime using runtime checks,
is said to be a \emph{safe} language.

This paper introduces the idea of making a distinction between safe
and unsafe simulations. A \emph{simulation} is said to be \emph{safe}
if no untrapped simulation errors occur. A \emph{simulation
  environment} is said to be safe if no untrapped simulation errors
can occur in any simulation. As a consequence, a natural question is
then what we mean by \emph{simulation error}. This paper focuses on
two kinds of simulation errors that can occur in hybrid modeling
languages~\cite{ModelicaLatestSpec,Ptolemaeus:2014,BromanSiek:2012,Nilsson:2003a}
and cosimulation environments~\cite{BromanEtAlFMI:2015,Cremona:2017}.  More
specifically, this work concerns both error classification and
solution methods. It presents the following main
contributions\footnote{All examples in the paper are available here:
  \url{http://www.modelyze.org/limbo}}:
\begin{itemize}
\item The paper describes two kinds of simulation errors that have
  traditionally been seen as modeling errors and not as untrapped
  simulation errors. More specifically, the errors concern i)
  \emph{unsafe zero-crossing detection}, and ii) \emph{unsafe
    accidental determinism} (Section~\ref{sec:safetyproblems}).
\item It describes an approach to make these untrapped simulation
  errors trapped, by introducing the concept of \emph{limbo} state. A
  simulation enters the limbo state when a simulation error is
  detected. The modeler has the choice of defining the behavior
  to leave the limbo state in a safe way and continue the simulation,
  or to terminate the simulation and report the error as a trapped error
  (Section~\ref{sec:limbo}).
\end{itemize}

\section{Hybrid Simulation Safety Problems}
\label{sec:safetyproblems}
This section describes two problems with hybrid simulation
safety. First, it discusses the infamous bouncing-ball problem,
where the numerical accuracies of standard zero-crossing detectors make
a bouncing ball to tunnel through the ground. Second, the section
discusses the relations between \emph{accidental} and
\emph{intentional nondeterminism}, and the safety problem resulting
from \emph{accidental determinism}. The latter problem is illustrated by
simultaneous elastic collisions of frictionless balls.

\subsection{Unsafe Zero-Crossing Detection}
\label{sec:unsafezero}

One classic simple example for demonstrating hybrid modeling and
simulation is the bouncing ball model. The model demonstrates how a
ball is falling towards the ground, and bounces with an inelastic
collision, thus bouncing with decreased height. This model can be
expressed in any modeling language that supports i) a continuous
domain for expressing velocity and acceleration, ii) a construct to
numerically detect the collision, and iii) an action statement that
changes the sign and magnitude of the velocity of the ball. The
following model is a straight forward implementation in the
Modelica language:

\begin{lstlisting}[language={Modelica},numbers=left]
model BouncingBall
  Real h,v;
  parameter Real c = 0.7;
initial equation
  h = 3.0;
equation
  der(h) = v;
  der(v) = -9.81;
  when h <= 0 then
    reinit(v, -c*pre(v));
  end when; 
end BouncingBall;
\end{lstlisting}

\noindent The model is divided into three sections. The first section
(lines 2-3) defines the two state variables (\verb|h| for the height
of the ball and \verb|v| for the velocity), and one parameter \verb|c|
that states the fraction of the momentum that remains after a
collision with the ground. The second section (line 5) states an
initial equation. In this case, the height of the ball is initiated to
value $3$. Note that a Modelica tool will implicitly initialize the
other variables to zero, in this case the velocity \verb|v|. The third
section (lines 7-11) declaratively states the equations that holds
during the whole simulation. The \verb|der| operator denotes the
derivative of a variable. For instance, \verb|der(h)| is the
derivative of the height. Lines 9-11 lists a \verb|when| equation,
which is activated when the guard \verb|h <= 0| becomes true. That is,
when the ball touches the ground (\verb|h| becomes approximately $0$)
the \verb|reinit| statement is activated. The \verb|reinit| statement
reinitializes state variable \verb|v| to the value of expression
\verb|-c*pre(v)|, where \verb|pre(v)| is the left limit value of
\verb|v|, before impact. Note how the \verb|-c| coefficient both
changes the magnitude and the direction of the ball. Although the
bouncing ball example is often used as a ``hello world'' model for
hybrid modeling, it also demonstrates two surprising effects.

\begin{figure*}[!b]
\center
\includegraphics[width=0.49\textwidth]{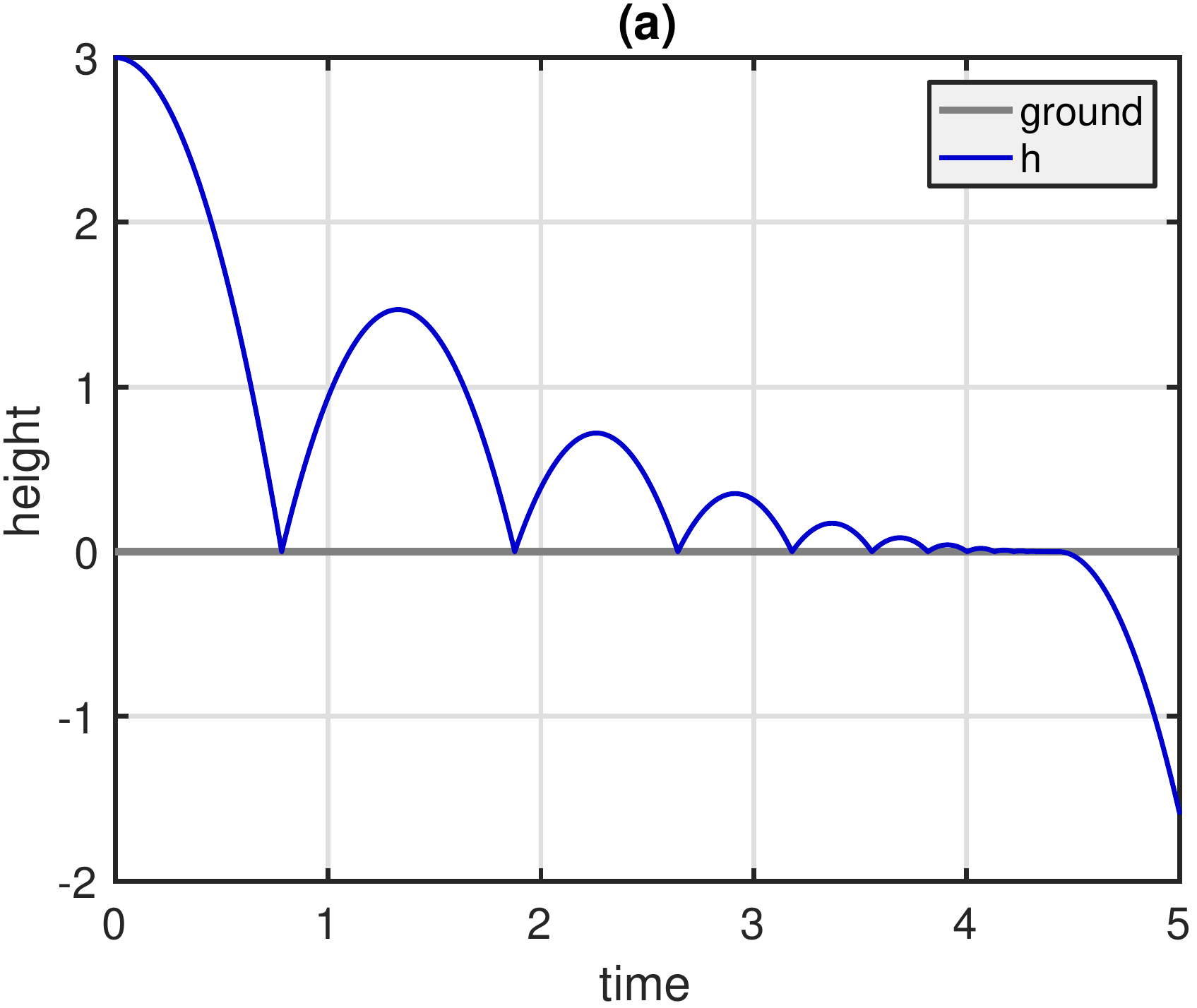}
\includegraphics[width=0.49\textwidth]{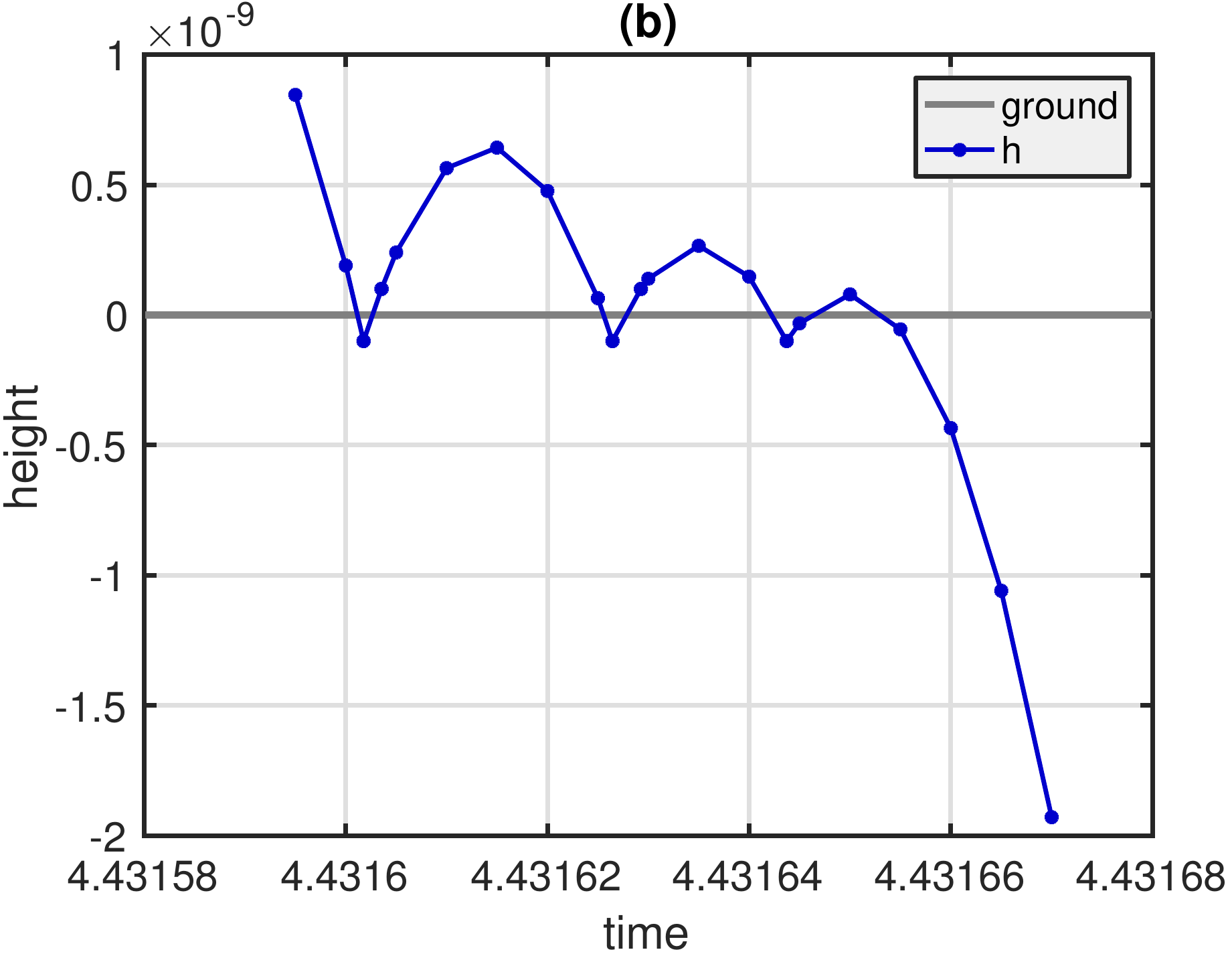}\includegraphics[width=0.49\textwidth]{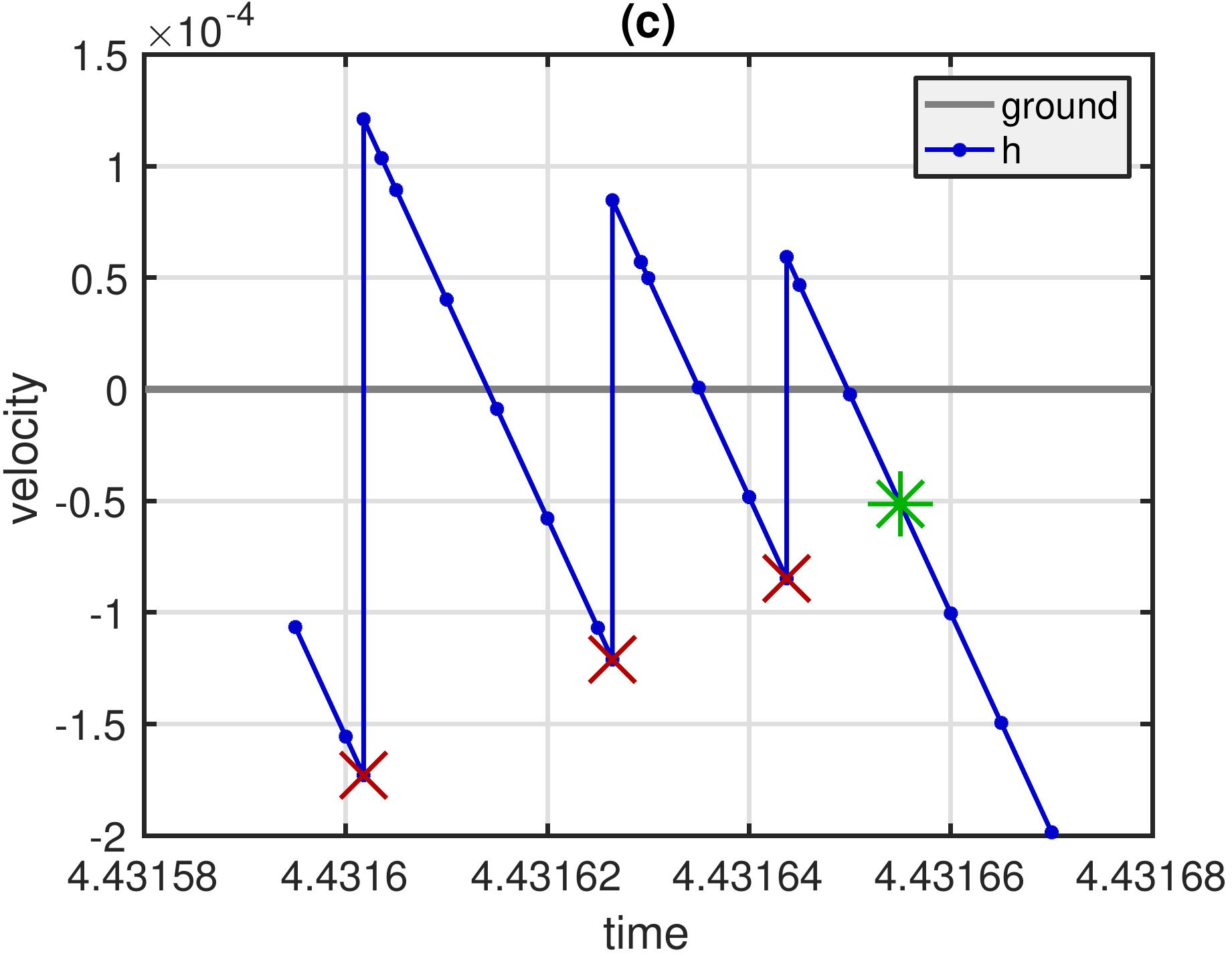}
\caption{A bouncing ball that is incorrectly tunneling through the
  ground. Figure (a) shows the height of the ball during the whole
  simulation, whereas the two bottom figures zoom in to the
  tunneling effect, showing the height (b) and the velocity
  (c). The simulation was performed using OpenModelica v1.11.}
\label{fig:bouncingball}
\end{figure*}

Fig.~\ref{fig:bouncingball}(a) shows the simulation result, plotting
the height of the ball. As expected, the ball bounces with decreased
altitude until it visually \emph{appears} to sit still, but then
suddenly tunnels through the ground. The model demonstrates two
phenomena. First, it shows an example of \emph{Zeno behavior}, where
infinite number of events (triggering the \verb|when| construct in
this case) in finite amount of time. The ball continuous to
bounce with lower and lower bounces. Second, the simulation trace
shows a tunneling effect, where the ball falls through the
ground. Fig.~\ref{fig:bouncingball}(b) shows the last bounces before the
tunneling effect. Note how the height of the last bounce is less than
$10^{-9}$ units.

Note, however, that the tunneling effect is not a consequence of the
Zeno condition, but of a numerical effect of how traditional
zero-crossing detectors detect and handle zero crossings. As can be
seen in Fig.~\ref{fig:bouncingball}(b), a zero-crossing detector
typically overshoots the crossing slightly, before the action is
applied. To be able to detect a crossing, the bounce needs to get over
a certain tolerance threshold, for the crossing to be
detected. Fig.~\ref{fig:bouncingball}(c) shows the velocity for the
last three bounces. The red crosses mark where zero crossings take
place, and where the velocity is changed from positive to negative in
the same time instance.
At the last instance (marked with a
green star) a zero crossing should have occurred, but the bounce has
not reached over the tolerance level above zero. Hence, no zero
crossing occurs, and the ball tunnels through the ground.

It is important to stress that the Zeno behavior
of the model and the numerical tunneling problem are two different
things. The former is a property of the model, whereas the latter is a
simulation error due to numerical imprecision in a specific simulation
tool. The Zeno effect has been extensively studied in the area of
hybrid automata, where regularization techniques are used to solve the
problem by creating a new model~\cite{JohanssonEtAl:1999}. Focus is on the zeno behavior and
not the tunneling problem.  Traditionally, also the tunneling effect has
been seen as a model problem. However, this paper argues the
opposite. The tunneling problem is a consequence of an \emph{untrapped
  simulation error}. A safe simulation environment should handle such
problems as \emph{trapped errors}, either by generating an exception
state that can be handled within the model, or by terminating the
simulation and report an error, before the tunneling effect occurs. A
potential solution is discussed in Section~\ref{sec:limbo}.

\subsection{Unsafe Accidental Determinism}
\label{sec:unsafedet}

The second problem has been extensively discussed in two
recent papers by Lee~\cite{Lee:2014,Lee:2016}. In these papers, Lee
discusses the problem of deterministic behavior of simultaneous
events, and illustrates the problem using an example with three
colliding balls. This section discusses the problem with the same
example, but using Modelica instead of Ptolemy II. The key insight in
this section is not the difference in modeling environment, but to
view the problem as a simulation safety problem, rather than a
modeling problem.

\begin{figure*}[!b]
  \center
\includegraphics[width=0.49\textwidth]{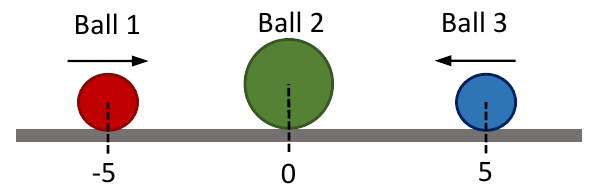}
\caption{Illustration of the example with three colliding balls. The
  balls roll without any friction. Ball 1 and ball 3 move with constant
  speed, where as ball 2 is sitting still before the impact. Note that the problem
  is a 1-dimensional problem: the balls can only move horizontally, and not in vertical
  directions.}
\label{fig:balls}
\end{figure*}

Consider the example in Fig.~\ref{fig:balls} where ball 1 and ball 3
are moving towards ball 2, which is sitting still. In the example, we
assume a frictionless surface and perfectly elastic collision, that
is, no energy is lost when the balls collide.

\noindent The following Modelica model defines the dynamics of a frictionless
elastic ball, with two state variables: \verb|x| for the horizontal position,
and \verb|v| for the velocity.

\begin{Modelica}
model Ball
  Real x;              // Position state
  Real v;              // Velocity state
  parameter Real x0;   // Initial position
  parameter Real v0;   // Initial velocity
  parameter Real m;    // Mass of the ball
  parameter Real r;    // Radius
initial equation
  x = x0;
  v = v0;
equation
  der(x) = v;     // Relation between position and speed  
  der(v) = 0;     // Constant speed, no acceleration
end Ball;
\end{Modelica}

\noindent For an elastic collision, the
momentum and the kinetic energy are preserved.
\begin{align}
m_1 v_1 + m_2 v_2 = m_1 v'_1 + m_2 v'_2, \qquad
\frac{m_1 v_1^2}{2} + \frac{m_2 v_2^2}{2} = \frac{m_1 {v'_1}^2}{2} + \frac{m_2 {v'_2}^2}{2}
\end{align}

\noindent Variables $v_1$ and $v_2$ represent the velocity before the collision for ball 1 and ball 2, respectively. The velocity after collision is given by $v'_1$ and $v'_2$. We then have 

%
%

\begin{align}
  v'_1 = \frac{2m_2v_2 + m_1 v_1 - m_2 v_1}{m_1 + m_2}, \qquad
  v'_2 = \frac{2m_1v_1 + m_2 v_2 - m_1 v_2}{m_1 + m_2}
  \label{eq:v}
\end{align}

\noindent in the case where $m_1 \neq 0$ and $m_2 \neq 0$. By
instantiating the \verb|Ball| model into three
components \verb|b1|, \verb|b2|, and \verb|b3|, we get the following model:

\begin{Modelica}
model ThreeBalls
  Ball b1(x0=-5, v0= 1, r=0.5, m=1);
  Ball b2(x0= 0, v0= 0, r=1.0, m=2);
  Ball b3(x0= 5, v0=-1, r=0.5, m=1);
equation
  //Detecting collision between ball 1 and ball 2
  when b2.x - b1.x <= b1.r + b2.r then  
    reinit(b1.v, (2*b2.m*pre(b2.v) + b1.m*pre(b1.v) - 
	       b2.m*pre(b1.v))/(b1.m + b2.m));
    reinit(b2.v, (2*b1.m*pre(b1.v) + b2.m*pre(b2.v) - 
	       b1.m*pre(b2.v))/(b1.m + b2.m));
  end when;
  //Detecting collision between ball 2 and ball 3
  when b3.x - b2.x <= b2.r + b3.r then  
    reinit(b2.v, (2*b3.m*pre(b3.v) + b2.m*pre(b2.v) - 
	       b3.m*pre(b2.v))/(b2.m + b3.m));
    reinit(b3.v, (2*b2.m*pre(b2.v) + b3.m*pre(b3.v) - 
	       b2.m*pre(b3.v))/(b2.m + b3.m));
  end when;
end ThreeBalls;
\end{Modelica}

\noindent Note that components \verb|b1| (ball 1) and \verb|b3| (ball
3) have the same mass \verb|m=1| and radius \verb|r=0.5|, whereas
\verb|b2| (ball 2) has \verb|m=2| and \verb|r=1.0|. The start positions are
$-5$, $0$, and $5$, for balls 1, 2, and 3, respectively.

\begin{figure*}[!b]
  \center
\includegraphics[width=0.49\textwidth]{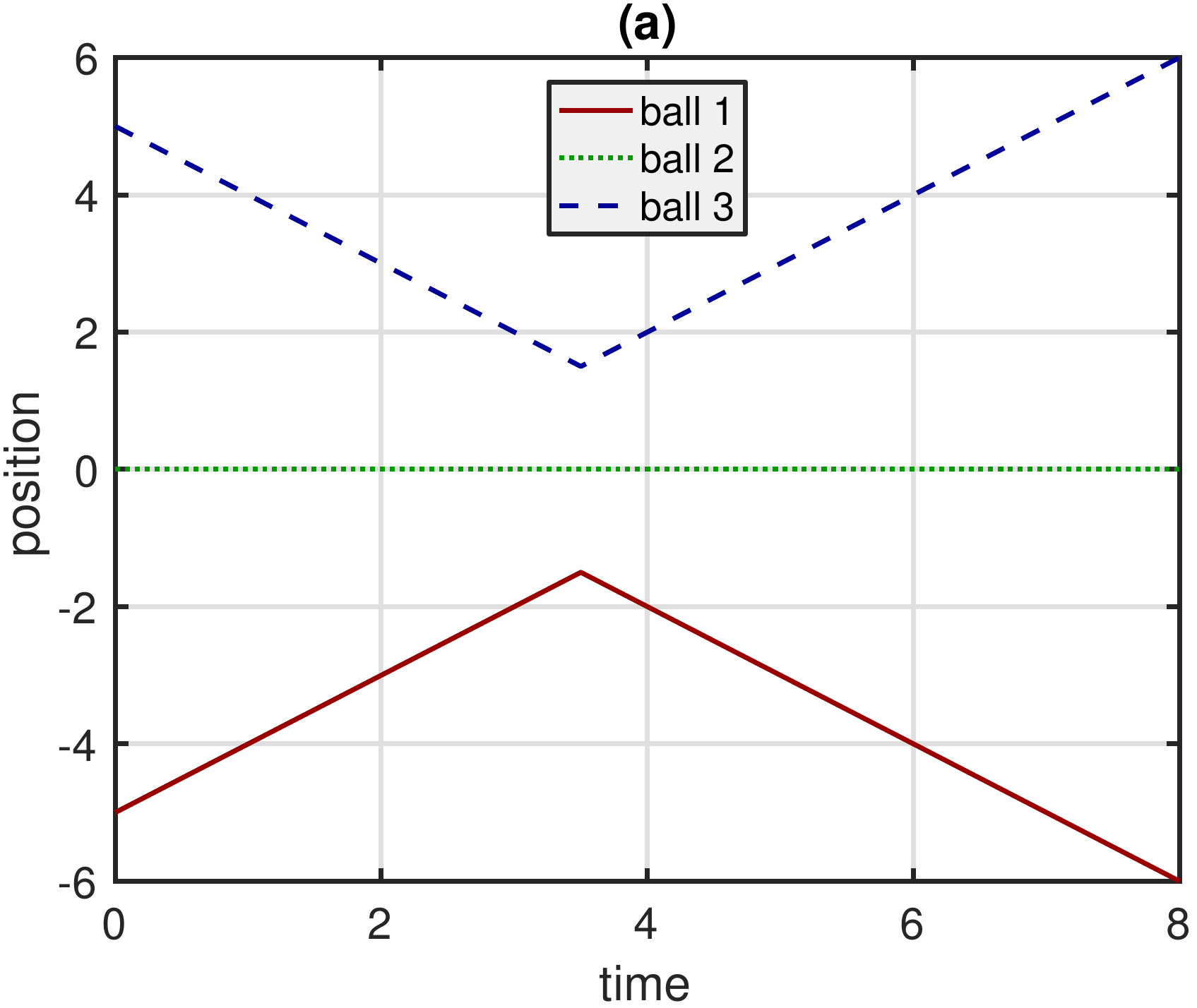}
~\\~\\
\includegraphics[width=0.49\textwidth]{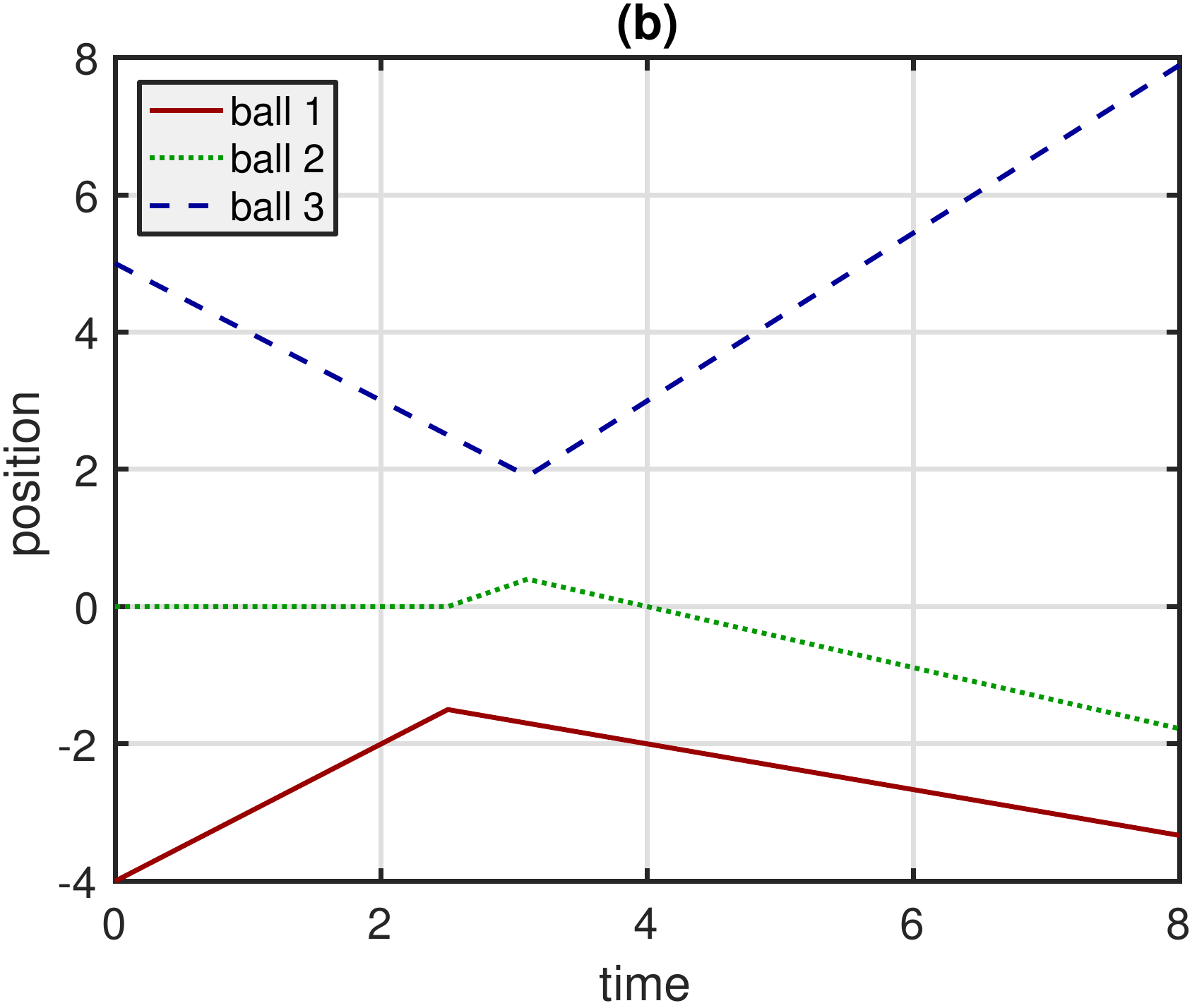}\includegraphics[width=0.49\textwidth]{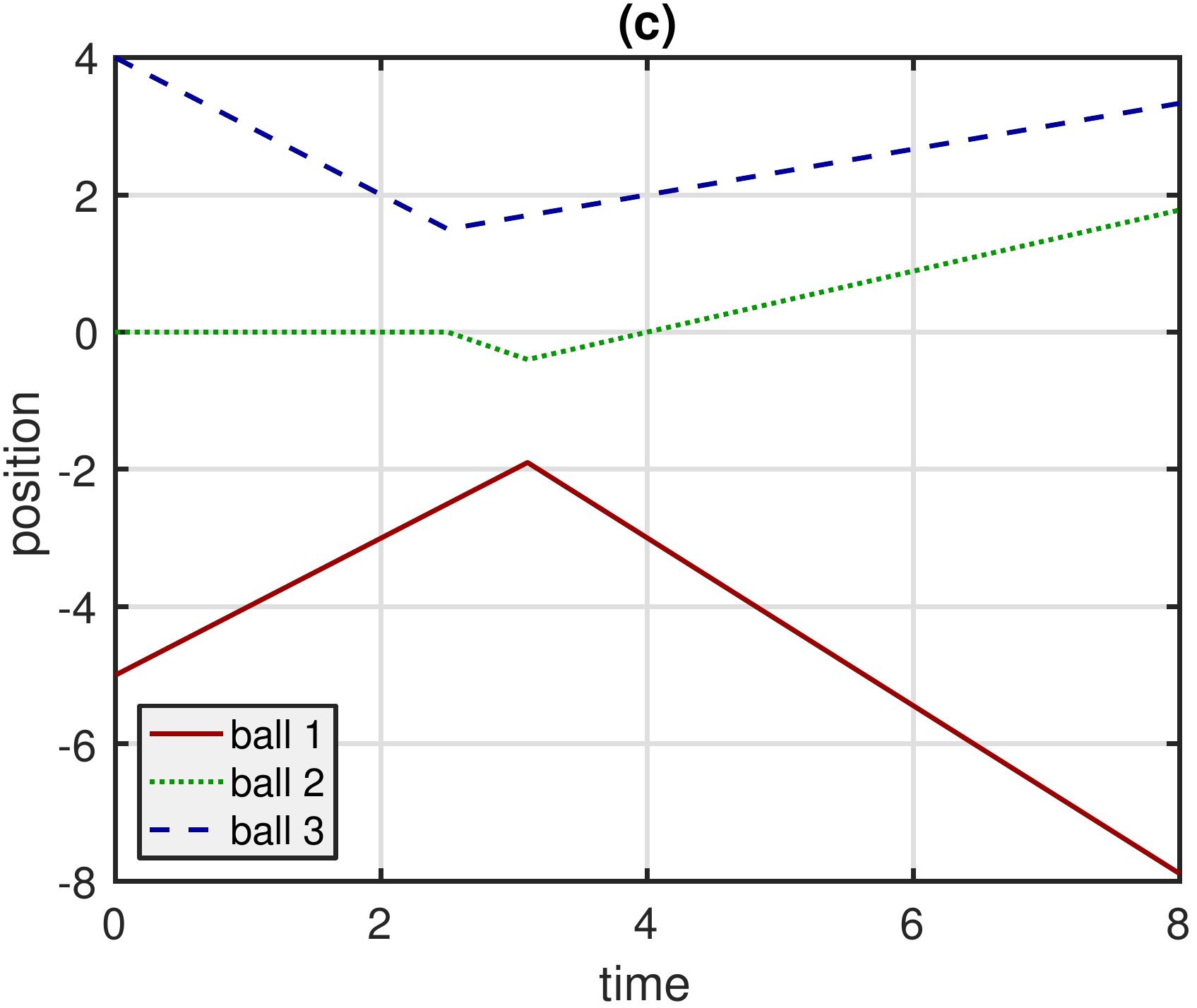}
~\\~\\
\noindent\includegraphics[width=0.49\textwidth]{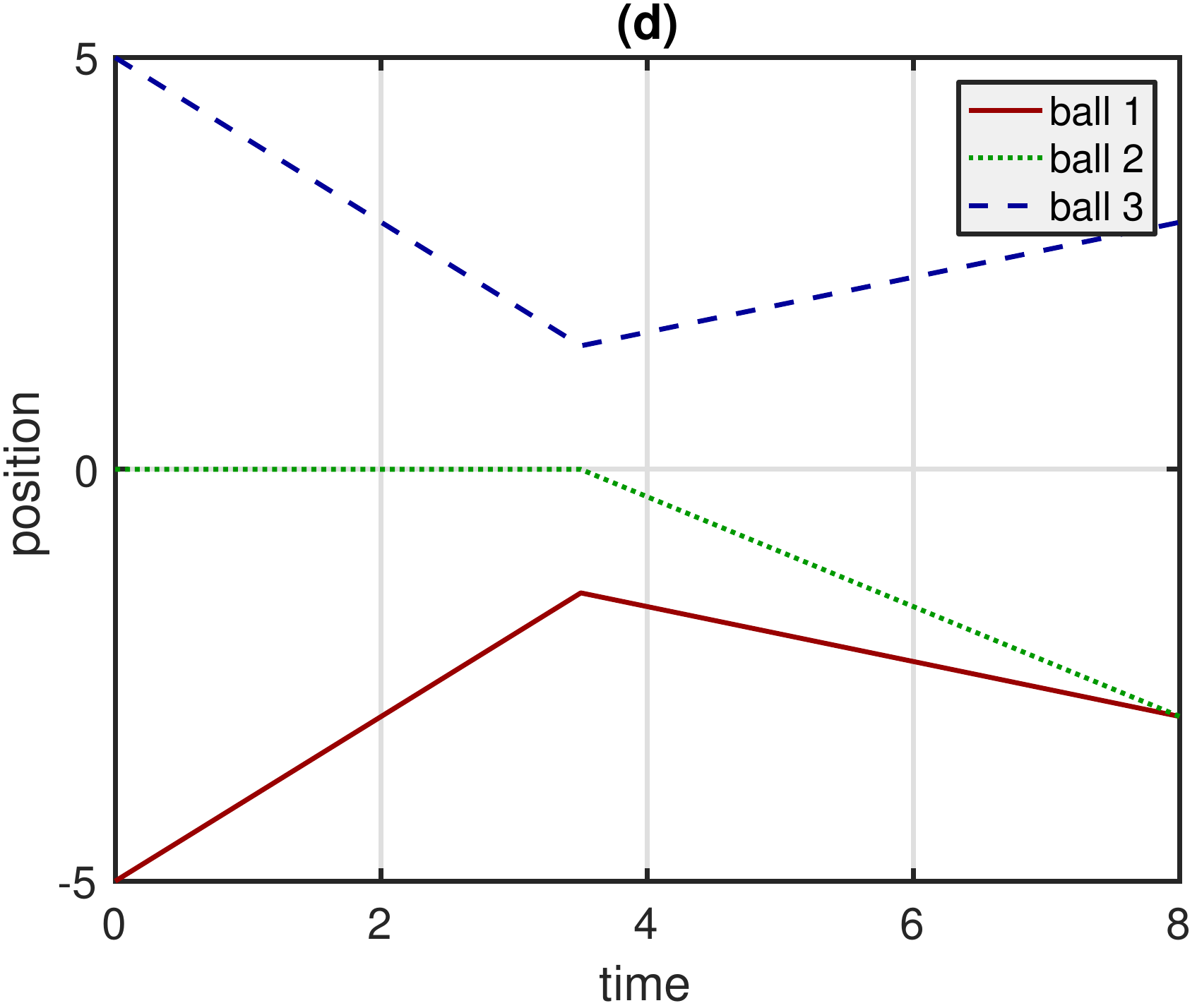}\includegraphics[width=0.49\textwidth]{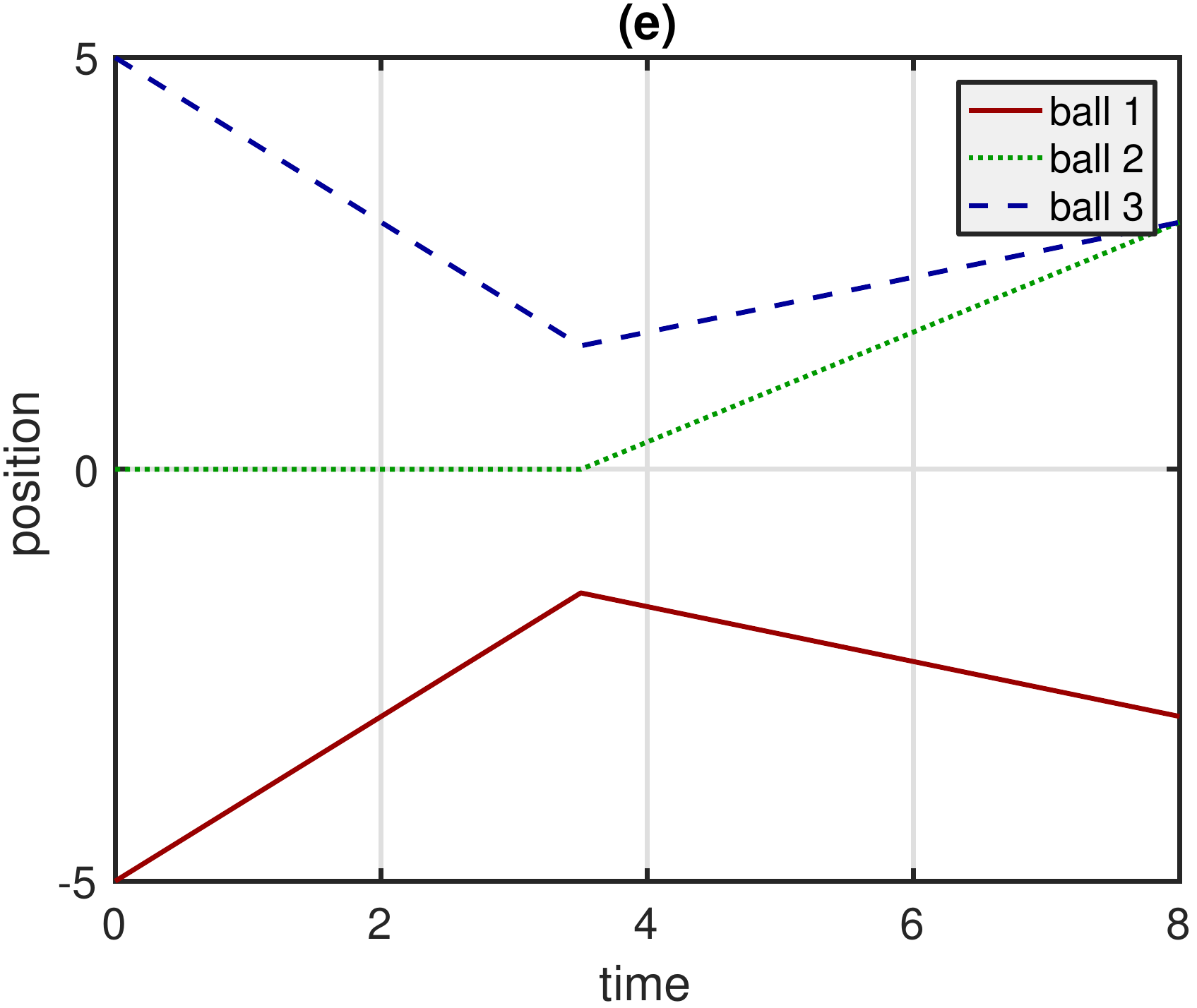}
\caption{Simulation cases for the $\texttt{ThreeBalls}$ model. Figure
  (a) shows an ideal result when the balls have the same initial
  distances. Figure (b) and (c) show simulation traces, where
  initial distances between the balls are not equal. Figures (d) and (e) show
  unexpected simulation results, where the initial distances between the
  balls are the same.}
\label{fig:collidingballs}
\end{figure*}

The changes in velocity, according to equations (\ref{eq:v}), are encoded
as two \verb|when| equations, each detecting either the collision
between ball 1 and 2, or between ball 2 and 3. What is then the
expected simulation trace for this model? One expected output might be
the plot in Fig~\ref{fig:collidingballs}(a). That is, a simultaneous
collision occurs, ball 2 does not move, and the other two balls bounce
back with the same velocity. This is actually \emph{not} what
happens when simulating the model. Let us take a step back and study the behavior of this model
in more detail.

Consider Fig~\ref{fig:collidingballs}(b) and
Fig~\ref{fig:collidingballs}(c). These two simulation traces show the
same example as above, with the difference that in
Fig~\ref{fig:collidingballs}(b), ball 1 starts a bit closer to the
middle ball, whereas in Fig~\ref{fig:collidingballs}(c) ball 3 starts
a bit closer.  As expected, in the first case, ball 1 hits ball 2
first, that makes ball 1 bounce back (it is the lighter of the two)
and ball 2 starts to move towards ball 3. Then ball 2 hits ball 3,
which bounces back and ball 2 changes direction again. In the first
case, ball 3 bounces back at a higher speed because the middle ball's
energy from the first hit gives ball 3 the extra speed. As expected,
Fig~\ref{fig:collidingballs}(c) shows the reverse, when ball 3 hits
ball 2 first.

Now, imagine that the distances between ball 1 and the middle ball,
and ball 3 and the middle ball become closer and closer to equal. As
long as one of the balls hits first, this will affect the other
ball. Hence, the limit for the two cases are not the same. As
Lee~\cite{Lee:2016} points out, the model may be seen as
nondeterministic in the case when both the balls collide
simultaneously. In that case, either ball 1 or ball 3 hits the middle
ball first, but nothing in the model indicates the order. Again, the
model is nondeterministic in this specific point.

In the previous two plots, the distances between the balls were not
the same. Fig~\ref{fig:collidingballs}(d) shows the actual simulation
result when simulating \verb|ThreeBalls| where the distances between
the balls are equal. We get a simulation trace, but is it the correct
one? Obviously no. We can notice two things. First, even if ball 1 and
2 arrives at the same speed from the same distance to the middle ball,
ball 2 moves to the left after impact. Why is the ball moving in that
direction and not the opposite direction? Second, note how ball 1 and
2 tunnel through each other, and are at the same position at time 8
(which should be physically impossible). The reason ball 2
moves to the left is that both \verb|when| equations are activated
simultaneously and that the code within the two \verb|when| blocks (lines
8-11 and lines 15-18) are executed in the order that they are stated
in the model. Hence, the velocity for ball 2 is initialized twice
(lines 10 and 15), where the last one (line 15) gives the final
result.

To make the situation even worse, assume that we switch the order of
the two \verb|when| equations in model \verb|ThreeBalls|, that is, the
\verb|when| equation for detecting collisions between ball 2 and 3
comes before the \verb|when| equation for detecting collisions between
ball 1 and 2. Modelica is a declarative language, where the order of
equations should not matter. Hence, we might expect to get the same
incorrect result. Unfortunately, this is not the case. The simulation
result for the new model, where the \verb|when| equations have
switched order, is shown in Fig~\ref{fig:collidingballs}(e). Note how
ball 2 moves in the opposite direction after impact because the order
of the impact from ball 1 and 3 has changed. 

Lee~\cite{Lee:2016} argues, based on a similar example, as follows:
``It would probably be wise to assume that determinism is incomplete
for any modeling framework that is rich enough to help design an
understand CPS, where discrete and continuous behaviors inevitably
mix.''. If the order of the evaluation of \verb|when| equations matter
and the order is left unspecified, the model is indeed
nondeterministic: there are two possible interpretations. However,
this paper argues that it is important to not mix the two separate
issues of the determinism of the model, and the determinism of the
simulation.

\begin{figure*}[!b]
  \center \includegraphics[width=0.8\textwidth]{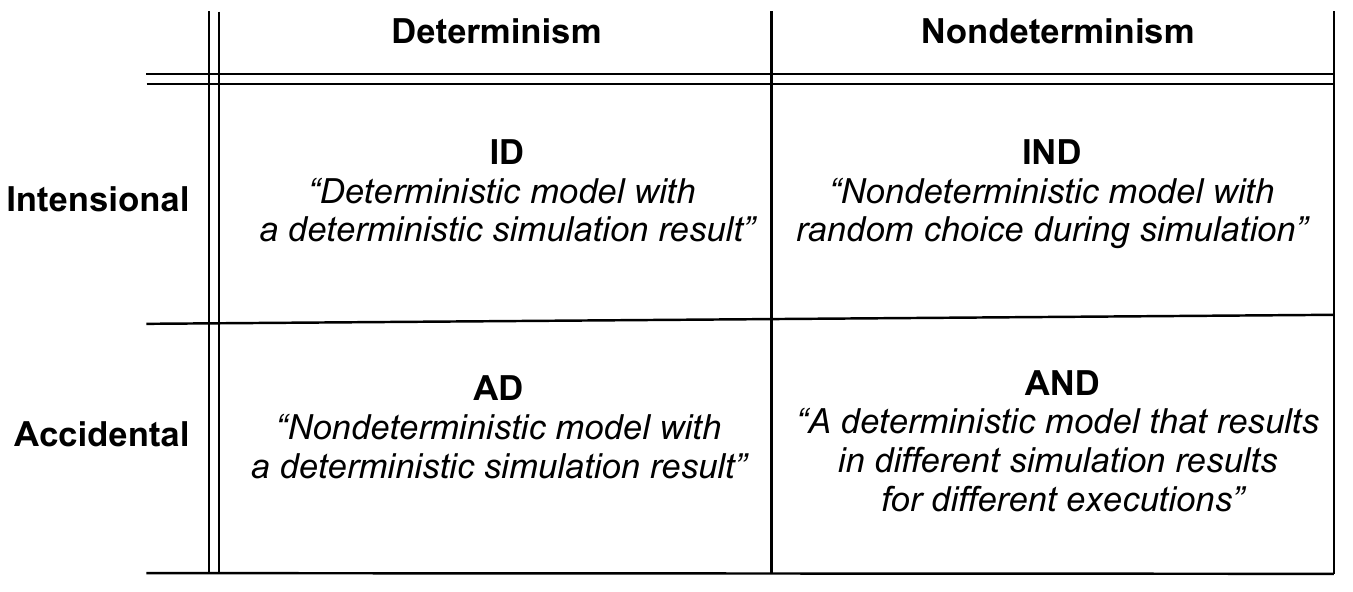}
\vspace{-2mm}
\caption{A matrix that shows the relationship between intensional
  determinism/nondeterminism and accidental
  determinism/nondeterminism. }
\label{fig:matrix}
\end{figure*}

Fig.~\ref{fig:matrix} shows a matrix, where we introduce the
concepts of intensional determinism/nondeterminism, and accidental
determinism/nondeterminism. \emph{Intensional determinism (ID)} for a
modeling and simulation environment is typically what is intended in
many simulation environments for cyber-physical systems. ID
means that the simulation of deterministic models yields deterministic
simulation results. The same model simulated with the same input
always results in the same simulation result. \emph{Intensional
  nondeterminism (IND)} means that the model itself is
nondeterministic, and that the simulator may use random samples to
generate the simulation result. Monte Carlo methods fall within this
category. Many useful formalisms, languages, and environments fall
within the categories of ID and IND.

The accidental categories are more problematic. \emph{Accidental
  nondeterminism (AND)} is when a simulator for a deterministic model
generates different simulation traces, even if the same model with the
same input is used. If a simulator behaves within the AND-category, it
typically means that there is an error in the simulator. For instance,
if a simulator is incorrectly using a multithreaded execution
environment, where the simulation result depends on the thread
interleaving, the simulator might give different results for different
executions.

The last category, \emph{accidental determinism (AD)} is the one that
is particular interesting in this example. In this case, a
nondeterministic model always yields the same simulation result. This
is exactly what happens in our simulation example of the
\verb|ThreeBalls| model. From Fig~\ref{fig:collidingballs}(b) and
Fig~\ref{fig:collidingballs}(c) we know that the order in which balls
1 and 3 hit ball 2 has a direct implication on the simulation
result. When the distance between the balls is the same, both
\verb|when| equations are activated simultaneously. The Modelica tool
then decides on an evaluation order for the equations. If the
\verb|reinit| statements were independent of each other, the order
would not matter. However, in this case, the order matters. As it
turns out, the simulator (OpenModelica v1.11~\cite{FritzsonEtAl:2005})
executes the constructs in linear order, which is the reason for the
different simulation traces for Fig~\ref{fig:collidingballs}(d) and
Fig~\ref{fig:collidingballs}(e). We have an accidental deterministic
behavior, where the original model was nondeterministic, but where the
simulation result is deterministic. Recall that the actual
activation choice is made on the order the \verb|when| equations are
defined in the file. Accidental determinism is an example of unsafe
simulation: the error is untrapped, that is, we get a simulation
result without warnings, even though the result itself is not deterministic.

\section{Safe Simulations using the Limbo State}
\label{sec:limbo}

The previous section showed two examples of unsafe simulation
behavior. In both cases, the simulation continued and produced a
result, without giving any errors or warnings. These are examples of
untrapped simulation errors. Although an error occurs at a specific
point in time (the tunneling effect or incorrect collision), the
simulator still produces a simulation result. The purpose of this
section is to illustrate the idea of how to make the untrapped errors
trapped, thus enabling safe simulations.

\subsection{The Limbo State}
The key idea is to introduce three conceptual states in a simulator:
i) the \emph{safe} state, ii) the \emph{limbo} state, and iii) the
\emph{unsafe} state. During simulation, the simulator is in one of
these three states. Note that these are states of \emph{the simulator}
itself, and not modes in a specific model. The idea of the limbo state
is first described abstractly, followed by a concrete discussion in
the context of the previous two problem examples.

\begin{figure*}[!b]
\begin{center}
\usetikzlibrary{positioning,automata}
\begin{tikzpicture}[->,>=stealth,shorten >=1pt,node distance=3cm,on grid]
  \node[state,initial]   (q_0)                {$\mathit{~~safe~~}$};
  \node[state]           (q_1) [right=of q_0] {$\mathit{~limbo~}$};
  \node[state,accepting] (q_2) [right of=q_1] {$\mathit{unsafe}$};
  \path[->] (q_0) edge [loop above]   node         {$a$} ()
                  edge [bend left=45] node [above] {$b$} (q_1)
            (q_1) edge                node [above] {$d$} (q_2)
                  edge [bend left=45] node [above] {$c$} (q_0);
\end{tikzpicture}
\end{center}
\vspace{-4mm}
\caption{A finite state machine diagram that includes the limbo state.}
\label{fig:fsm}
\end{figure*}
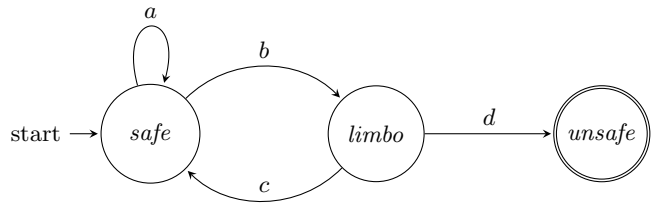

Fig.~\ref{fig:fsm} depicts a finite state diagram with the three
states. A simulation starts in the safe state. If no errors occur, the
simulator stays in the safe state. If a potential error occurs,
transition $b$ is taken to the limbo state. The limbo state means that
the simulator is in between a safe and an unsafe state. The error
\emph{is about to happen}, but has not yet taken place. From the limbo
state, either the simulation is safely terminated with an error
message (a trapped error), or transition $c$ is taken back to the safe
state. It is the modeler's responsibility to augment the model, such
that transition $c$ can be taken. If the error occurs in the limbo
state, transition $d$ is taken. If the simulator is safe, transition
$d$ should happen \emph{when the error occurs}, that is, it should
terminate the simulation at the simulation time of the error. Thus,
transition $d$ should generate a trapped error, indicating that the
simulation reached an unsafe state at a specific point in time.

\begin{figure*}[!b]
\center
\includegraphics[width=0.49\textwidth]{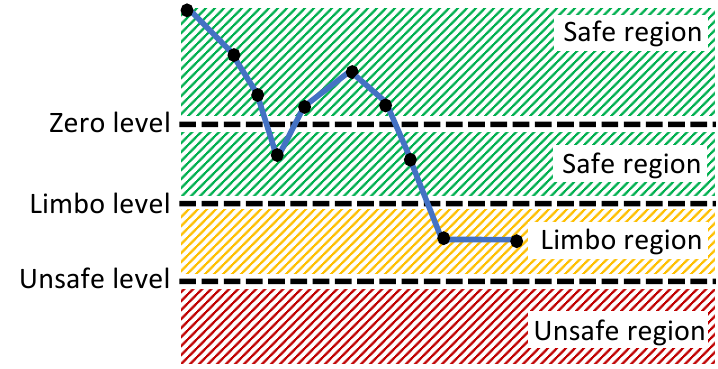}
\caption{The three different crossing detection levels and regions.}
\label{fig:levels}
\end{figure*}

The reader might now ask why we see the described problems as 
simulation errors, when the user still can modify the model to avoid
the error? Is it not a modeling error then? The point is, again, that
the error that appears during simulation must be trapped.
However, the same model can still be valid for
different simulation input. For instance, the bouncing ball model in
Figure~\ref{fig:bouncingball}(a) is valid before time 4, since the
simulation error happens sometime between time 4 and 5.
Let us now consider the two problems in
Section~\ref{sec:safetyproblems} in turn.

\subsection{Safe Zero-Crossing Detector}
The tunneling problem described in Section~\ref{sec:unsafezero} can
easily be detected using multiple levels of
zero-crossings~\cite{Tiller:2017}. The problem with traditional
zero-crossing detectors, such as the \verb|when| equations in
Modelica and level-crossing detector actors in Ptolemy II, is that
they can easily be used in an unsafe way. The key idea is instead that
a modeling language should \emph{only} provide safe zero-crossing
detectors, where the tunneling effect cannot occur.

Fig.~\ref{fig:levels} depicts the structure of a safe zero-crossing
detector. A safe zero-crossing detector has a safe region, a limbo
region, and an unsafe region.  The detector consists of three levels
of detection mechanisms: i) \emph{zero level} that detects the actual
zero crossing, ii) \emph{limbo level} that detects when the limbo
region is entered, and iii) the \emph{unsafe level}, which detects
that the model did not leave the limbo state correctly.

Returning to the state machine in Fig.~\ref{fig:fsm}. Transition $a$
is taken each time the \emph{zero level} is crossed. That is, the
simulation is still safe, even if the zero level is crossed. In the
bouncing ball example, this happens every time the ball bounces
correctly (see the example trajectory line
in Fig.~\ref{fig:levels}). Transition $b$ is taken if the variable value
crosses the limbo level. In the bouncing ball example, this occurs
when the ball is starting to tunnel through the ground. Note that this
does not have to be an error. If the modeler detects the tunneling
effect, and then changes the mode of the ball to stay still (no
acceleration or velocity), the simulation changes state to be safe
again (transition $c$), or stays in the limbo region (see again the
example trajectory line in Fig.~\ref{fig:levels}). However, if the model
is incorrectly implemented, as in the example in
Section~\ref{sec:unsafezero}, the unsafe level will be crossed. In
such a case, the simulation environment should generate a trapped
error, by terminating the simulation and by reporting the simulation time of
the error.
A safe modeling and simulation language should only include safe
zero-crossing detectors as primitives, making it impossible to use
unsafe zero-crossing detection.
Consider now the following Modelica model.

\begin{figure*}[!t]
\center
\includegraphics[width=0.49\textwidth]{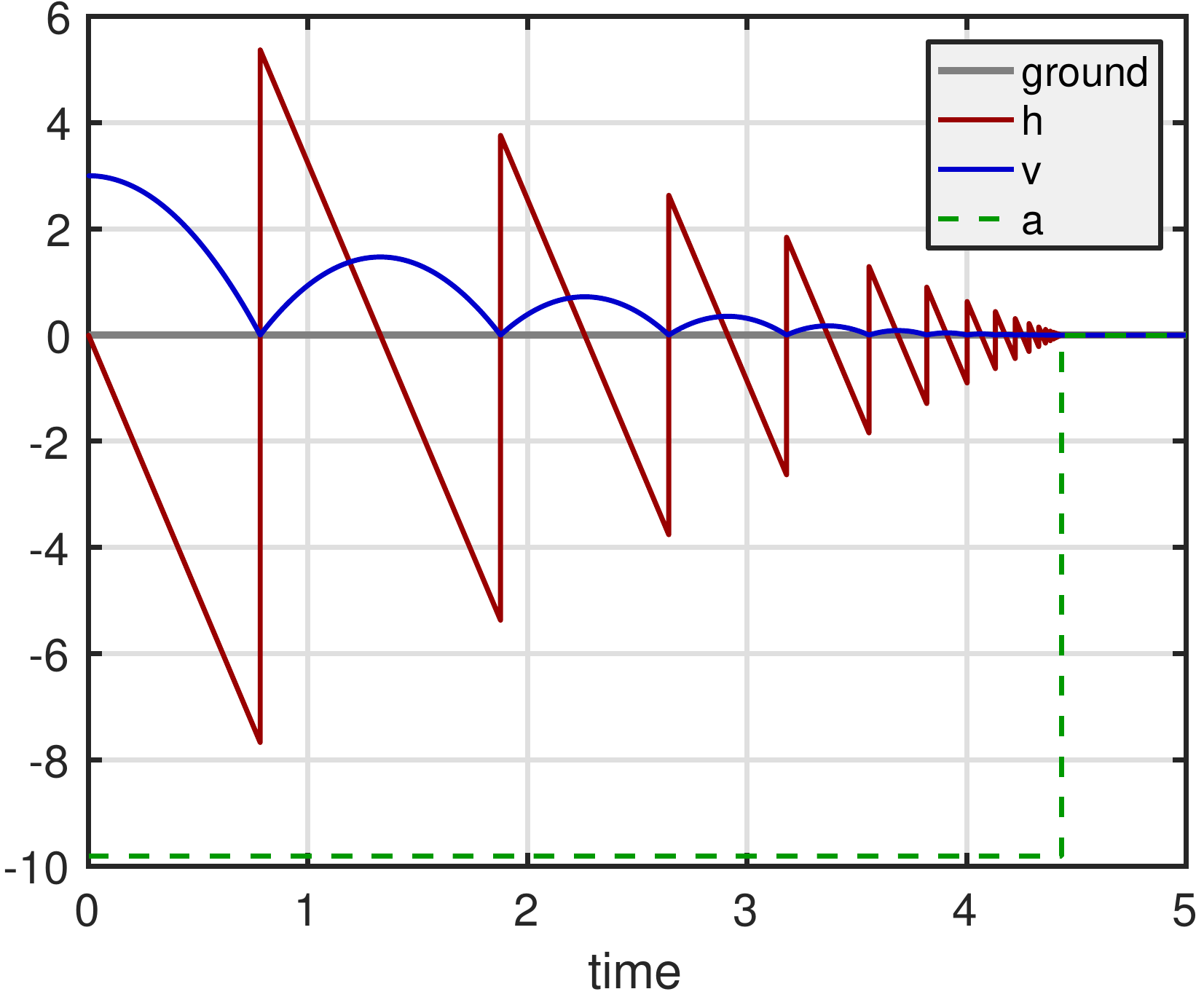}
\caption{A safe bouncing ball that stays on the ground. Note how the
  acceleration $\texttt{a}$ of the ball transitions from $-9.81$ to $0$
  when the ball comes to rest.}
\label{fig:safebouncingball}
\end{figure*}

\begin{lstlisting}[language={Modelica}]
model SafeBouncingBallFinal
  Real h,v;
  discrete Real a(start = -9.81);
  parameter Real c = 0.7;
  parameter Real epsilon = 1e-8;
  Boolean limbo;
initial equation
  h = 3.0;
  v = 0;
  limbo = false;
equation
  der(h) = v;
  der(v) = a;
  when h <= 0 then
    reinit(v, -c*pre(v));
  end when;
  
  //Limbo state action
  when limbo then
    reinit(v,0);
    a = 0;
  end when;  
  // Detecting limbo level
  when h <= -epsilon then
    limbo = true;
  end when;
  // Detecting unsafe level
  when h <= -2*epsilon then
    terminate("Unsafe Zero Crossing");
  end when; 
end SafeBouncingBallFinal;
\end{lstlisting}
~\\~\\
\noindent Fig.~\ref{fig:safebouncingball} shows the simulation trace
of simulating the model. A few remarks are worth making. We can see
that the \verb|when| equation for detecting the zero crossing is
unchanged compared to the previous section. What has been added are
two more \verb|when| equations that detect the limbo level (line 24),
and the unsafe level (line 28). If the limbo region is entered (line
24) a boolean variable is updated, which triggers the limbo action
(lines 19-22), where the ball is put to rest.  Note that if we reach
the unsafe region (line 29), the simulation is terminated. In this
case, this is done explicitly in the model, but an ideal modeling
language should include such detection automatically. It is not
obvious how to extend Modelica in this way, but an interesting
direction is to be able to specify invariants of safe states, as done with invariants
in hybrid automata~\cite{AlurEtAl:1993}.

\begin{figure*}[!t]
\center
\includegraphics[width=0.91\textwidth]{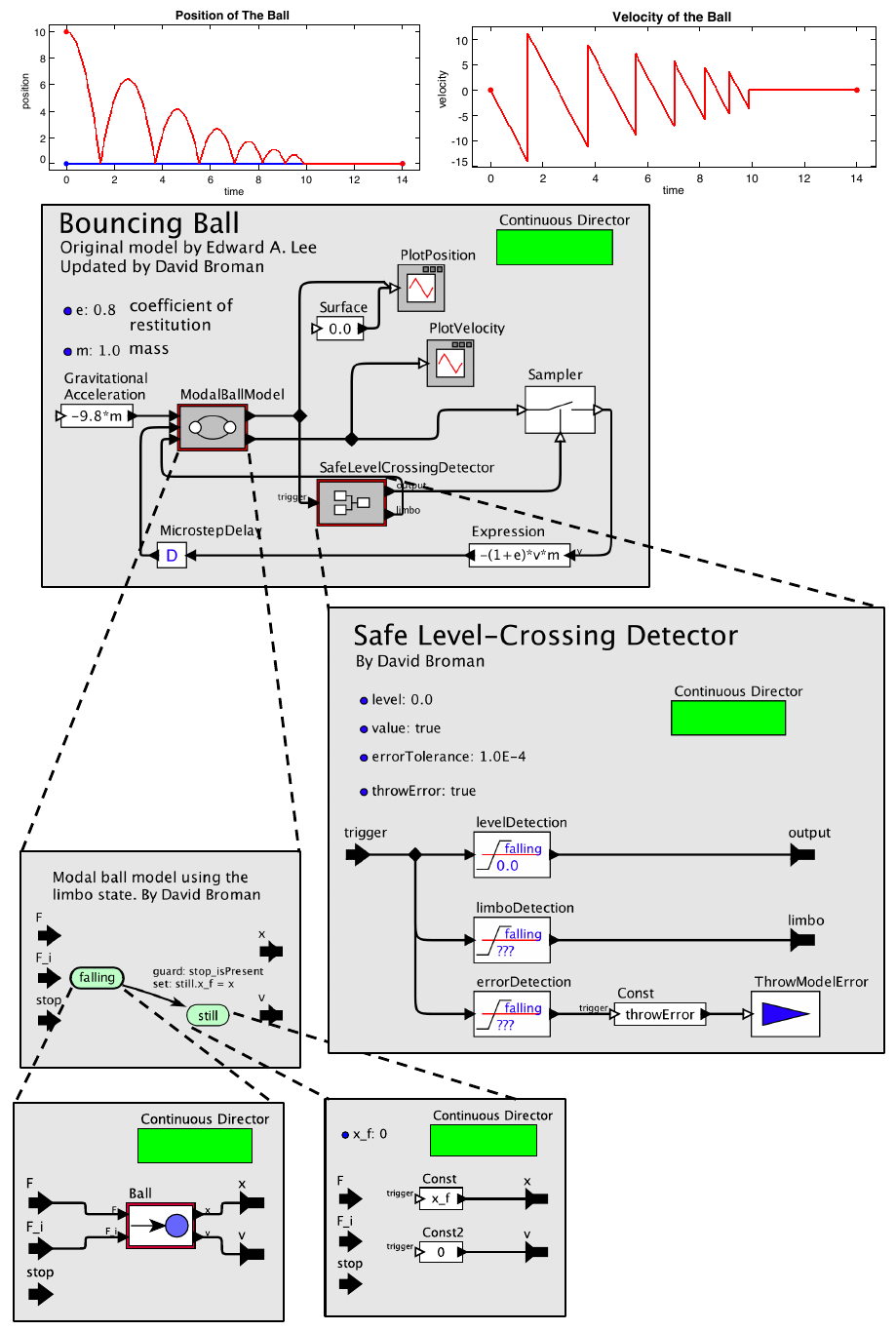}
\caption{An implementation of a safe level-crossing detector in
  Ptolemy II. The original open source model, before extending it with
  the safe level-crossing detector, is available
  here: $\texttt{http://ptolemy.org/constructive/models}$}.
\label{fig:ptolemy}
\end{figure*}

A zero-crossing detector can be generalized into a directional
level-crossing detector, that can detect arbitrary level in one
direction. Fig.~\ref{fig:ptolemy} shows a simple safe level-crossing
detector that is implemented as an actor in Ptolemy II. The main model
called \emph{Bouncing Ball} is a modified version of the bouncing ball
example from \cite{Lee:2014}. Two changes
have been made: i) the \verb|Ball| model has been replaced with a
\verb|ModalBallModel|, and ii) the original level-crossing detector has been
replaced with a new safe level-crossing detector. Note that the safe
level-crossing detector actor has approximately the same interface as
the level-crossing detector in the Ptolemy II standard library, with
the main difference that it also has an output port called
\verb|limbo|. The safe level-crossing detector outputs a discrete
event on the \verb|limbo| port if it detects a limbo state. When it is
used in the bouncing ball example, it means that the ball is just
about to start to tunnel. In the example model, the limbo port is
connected to the \verb|ModalBallModel| actor's \verb|stop| port. The
modal model has two modes, i) the ball is
\verb|falling|, and ii) the ball is sitting \verb|still|. If the modeler
forgets to connect the \verb|limbo| port, the safe level-crossing detector reports a trapped error.

\subsection{Safe Deterministic Event Handling}
In the colliding ball example in Section~\ref{sec:unsafedet}, the root
of the accidental deterministic behavior was simultaneous events. It
is extremely hard (if not impossible) to guarantee that events cannot
happen simultaneously. Numerical imprecision, both due to round-off
errors and integration errors, makes it hard to give any guarantees. A
naive solution would be to always enforce that no events occur
simultaneously by arbitrarily selecting an order. However, this will
lead to the problem of accidental determinism. If the order actually
matters, such arbitrary deterministic choice would result in an unsafe
behavior.

Instead, our proposal is to again make use of the limbo state diagram,
as shown in Fig.~\ref{fig:fsm}. The transition $b$ should be activated
when two events are sufficiently close to each other. The exact meaning of 
\emph{sufficiently close to each other} can be configured using a numerical
tolerance level.
This means that a model will transition into the limbo state when
simultaneous events occur. This does not have to be an error. If the
modeler knows how to handle the specific case, he/she can express this
in the model (assuming that the modeling language is expressive
enough) and then make a transition back to the safe state. If no such
case for simultaneous event is implemented, the simulation tool must
report a trapped error. In Modelica, \verb|elsewhen| constructs can be
used to implement such special cases. This is indeed what was done to
create the simulation plot in Fig~\ref{fig:collidingballs}(a).
Note that a nondeterministic model with missing cases can be seen as
an underspecified model. By adding all missing cases and completely
specifying the model, we convert a nondeterministic model into a
deterministic model.

\section{Conclusions}
\label{sec:conclusion}
This paper presents and discusses the idea of safe simulation. In
particular, it makes a distinction between trapped and untrapped
errors. As part of the solution, the notion of a limbo state is
introduced. The preliminary work is illustrated using small examples
in Modelica and Ptolemy II. However, to make the approach useful in
practice, the safety concepts need to be integrated as explicit parts
of a modeling language and a simulation environment. An interesting
direction for future work is to investigate if type systems in
modeling languages~\cite{BromanFritzsonFuric:2006} can be used to
statically detect and eliminate untrapped errors.

\section*{Acknowledgments}
I would like to thank Edward for the really great collaboration we
have had over the years. And, congratulations on your birthday! I would also
like to thank Cláudio Gomes and the anonymous reviewers for many
useful comments. Finally, I would like to acknowledge and thank Bernhard
Rumpe for pointing out the connection between nondeterminism in models
and underspecification.

This project is financially supported by the
Swedish Foundation for Strategic Research (FFL15-0032).

\bibliographystyle{plain}
\bibliography{references}

\end{document}